\documentclass{article} 
\usepackage{graphicx}
\begin{document}
\bibliographystyle{plain}
\title{Memorization in a neural network with adjustable transfer function 
and conditional gating}
\author{Gabriele Scheler}
\maketitle

\begin{abstract}
The main problem about replacing LTP as a memory mechanism has been 
to find other highly abstract, easily understandable principles for 
induced plasticity.
In this paper we attempt to lay out such a basic mechanism, namely 
intrinsic plasticity.
Important empirical observations with theoretical significance are  
time-layering of neural plasticity 
mediated by additional constraints to enter into later stages, 
various manifestations of intrinsic neural properties, and  
conditional gating of synaptic connections.
An important consequence of the proposed mechanism is that it can explain 
the usually latent nature of memories.
\end{abstract}

\section{Intrinsic Properties}

The intrinsic excitability of the neuron can be defined as an 
adjustable transfer function.
A neuron computes as a programmable element: it receives input that sets 
its transfer function and then computes inputs according to this transfer 
function until a new reset occurs.
An elementary transfer function may simply be constructed as a threshold 
logic where the threshold (number of elements necessary for producing an input)
changes. Another way to modify the transfer function is a bistable 
activation function as a modification of a sigmoidal activation. This 
form of modification has been reported specifically in striatal areas 
under dopamine modulation \cite{GruberAJetal2003}. 
Finally,
frequency-dependence of the transfer function has also been observed.
This corresponds to a U-shaped activation 
function with a range of input frequencies that activate a neuron the 
most \cite{HaasWhite2002,GarabedianCEetal2003}.

We may subject a number of isolated neurons with different intrinsic properties 
to a set of synaptic stimulations 
(e.g. low frequency, theta-like, high-frequency, irregular, regular)
and generate spiking responses.
The individual differences in the responses are the contributions from the 
intrinsic properties of the neuron.

\section{Encoding and Decoding by Intrinsic Plasticity}

Intrinsic plasticity is a potentially important concept, since it has 
often been 
observed that neurons code for features or modalities 
or events, 
such as place cells in rat and human hippocampus \cite{EkstromADetal2003}, or 
'category cells' in primate hippocampus \cite{HampsonREetal2004}.

When a neuron becomes activated in specific situations, it may acquire 
a characteristic combination of membrane proteins (receptors/ion channels) on 
its surface, and possibly characteristic 
localizations and concentrations of intracellular proteins.
Activated neurons may cluster together or form patches with 
specific (strong) connections among each other, but they may also be
more individual and localized (however, in this case the memory acquired may 
fail to last).
In this way, neurons acquire an identity.

\begin{figure}
\begin{center}
\includegraphics[width=6cm]{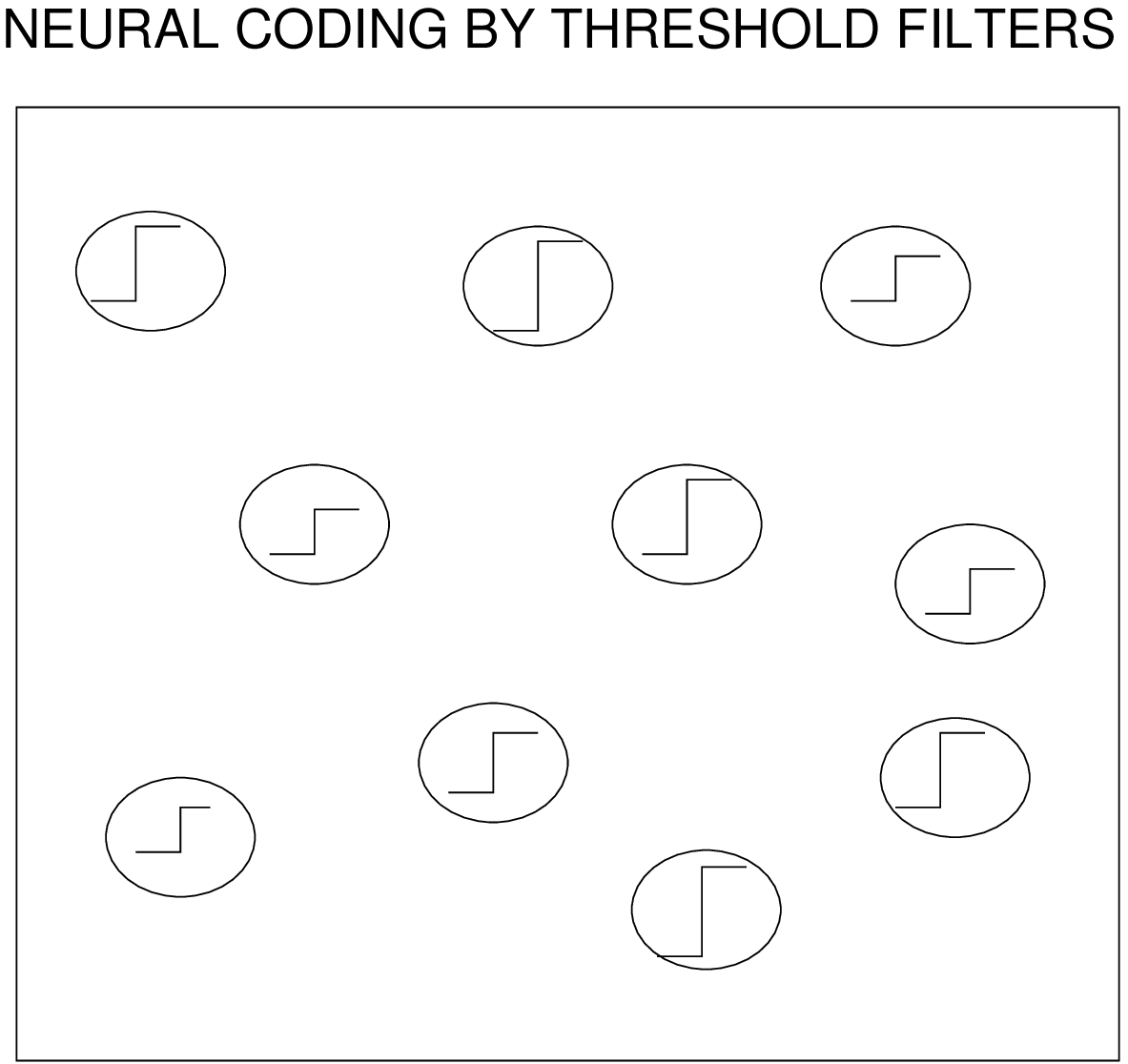}
\caption{Neurons with different activation 
thresholds (instantaneous firing rate)}
\label{threshold}
\end{center}
\end{figure}

When neurons have different thresholds of excitation, they will respond to 
activation by a concerted stimulus with different 
latencies (Fig.~\ref{threshold}). For early, fast responses to sensory 
stimuli, as in vision, the first-spike population response defines the basis 
for a sweep across neural hierarchies.
The population response is then 
partly defined by the specific strength with which stimulus-induced 
presynaptic firing acts on the neuron (synaptic strength), and partly 
defined by the respective threshold of activation that a neuron 
exhibits. 

\begin{figure}
\begin{center}
\includegraphics[width=6cm]{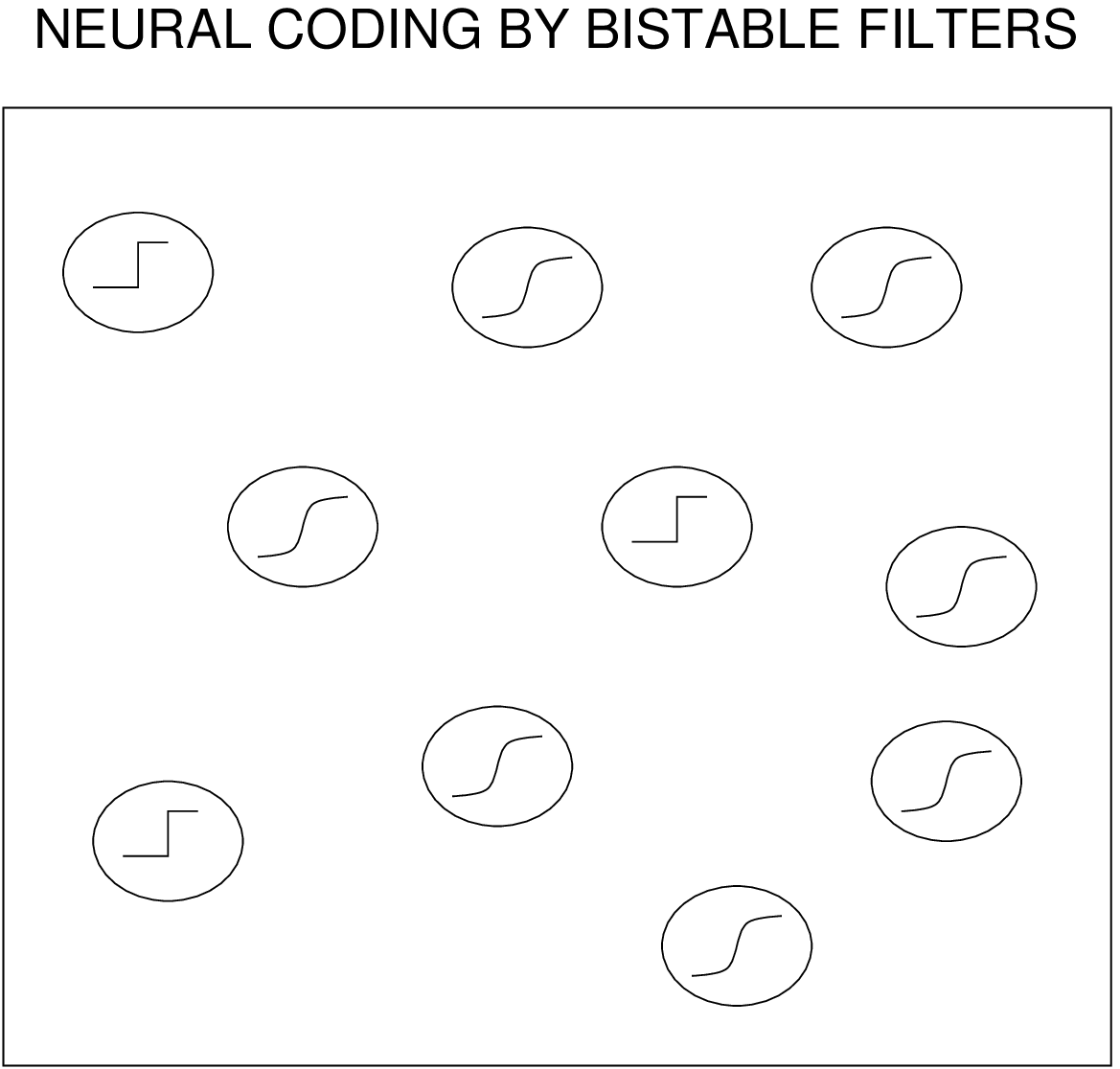}
\caption{Neurons with bistable modulation of their transfer function}
\label{bistable}
\end{center}
\end{figure}

Let us further assume that neurons for seconds and minutes may enter into a state 
where they respond with bistable activation to sustained input 
(ON or OFF firing patterns - binary version of the sigmoidal activation 
function), such as for dopamine modulation in striatum.
An example is given in Fig.~\ref{bistable}.

How may this be related to information storage and read-out?

Let's say we have 100 neurons coding 
for a motoric pattern such as an arm movement.
When I change the activation function in a number of these neurons, 
say 10\%, 
I may subtly alter the representational pattern that emerges.
The affected neurons will now fire high when activated above threshold,
and fire low, when not sufficiently activated. 
Non-affected neurons will respond with a linear graded output to an input.
Thus the firing pattern 
that is generated by a certain stimulus on those 100 neurons will 
be altered in 10 of them.

As an example, we may look at angles for a number of joints to define 
a movement.
A set of vectors between 0-1 may code for angles, such as 
between $5-12^\circ$.

A neuron with a sigmoidal activation function will respond with a 
graded response in the linear section, i.e. dependent on input it will 
produce an output corresponding to a value between 5 and $12^\circ$.
A neuron with a sharpened sigmoidal function, or a binary activation 
(bistable neuronal activation function) will correspond to a wide range 
of inputs essentially either by not responding, or by putting out a fixed 
value, here 12$^\circ$ (or 10-$12^\circ$).
In this sense, the neuron has acquired memory, it has acquired an informational 
value rather than a mere transfer function.
We may call this a case of coding by loss of information transfer and 
increase of storage.
This may happen for instance in  
skilled motor training, where an imprecise movement gets fine-tuned to 
a precise movement. 

\begin{figure}
\begin{center}
\includegraphics[width=6cm]{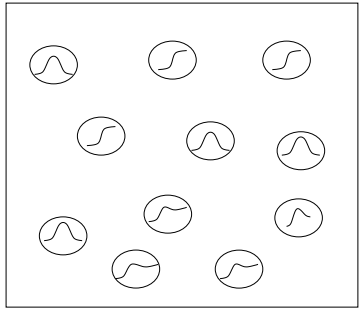}
\caption{Neurons with frequency-dependent transfer functions}
\label{frequency}
\end{center}
\end{figure}

A similar mechanism may underlie auditory cortex plasticity, except that 
this may be due to cholinergic modulation and involve a slight difference 
in the modulation of the activation function. In this case, the 
activation function may actually acquire a U-shaped curve for optimal 
frequency of input when it fires the most \cite{GarabedianCEetal2003},
cf. Fig.~\ref{frequency}.
A graded response neuron would give an output e.g. corresponding to  4-10 Khz,
while bandpass filtering may fixate it at 9 Khz.

As an example we can look at a neural map being stimulated by spike 
trains, sampled in 500 ms bins for their firing rate. Linear response 
neurons reproduce the input firing rate. Learned filter neurons fixate 
the output at their learned value, or produce a very small output. A 
combination of many linear and few filter neurons produces a neural map 
with individual neurons or areas of higher predictability and reduced 
input-dependence. This basic mechanism may be applied to purely technical 
problems, but this is outside the scope of this paper.

\section{Neurobiological background}

Neural filters can be activated by dopaminergic or cholinergic 
receptors.
To make these filters  permanent, ion channels such as inward rectifiers 
(GIRK) and high-voltage gated calcium channels may be upregulated
such that a neuron tends towards increased bistability even in 
the absence of neuromodulator input. The neuromodulator stimulation will 
then have the effect of switching the cell into a state, which, 
when continued, may migrate into an ion channel pattern. 

Our subjective experience of thoughts and images 
that follow each other may be a result of 
the memory that flickers in and out on the basis of seconds 
and minutes mediated by NM signaling. 

Ion channels can fixate the experience, which makes 
the memory independent of NM signaling.
For longer time-scales, 
gating of synaptic connetions (synaptic uncoupling/coupling) 
may be one possibility to store the information and making it only 
conditionally accessible.
Another possibility is transferring the memory even further from the 
membrane into intracellular changes.  This may involve a higher level 
of expression of a certain protein, with regulatory consequences, such that
even after the membrane protein expressions are statistically normalized,
brief triggering events lead to rapid or longer-lasting change of transfer 
function that do not occur in neurons that have not retained this form 
of intracellular information. Thus the neuron needs triggers to read out 
the information, but the same trigger that does not cause an alteration 
of membrane function in a naive neuron causes this alteration in an 
imprinted neuron. Candidate for such intracellular upregulation 
are PKC \cite{KohenRetal2003}, PKA \cite{ChaoNestler2004}, 
or RGS-proteins \cite{Burchett2003}, but there are many 
others \cite{WermeMetal2002,UslanerJetal2001}.

\section{Selective read-out}

How can information stored in individual neurons be read out and affect 
network processing?
Any activations by stimuli in a network where neurons 
exist with their different identities will activate a pattern over these 
neurons. It will depend on the new mixture of types of stimulations exactly
which pattern is being formed. 
One can then subtract the pattern that results from the synaptic input 
imposed on the network from the underlying pattern of stored neural identities.
The interactions between neurons complicate the patterns
since they overlay temporal activations at different time-scales.
But there is an important initial effect:
The closer the activation matches to the specific profile of a 
neuron - and this activation may involve several time-scales as well - the more this neuron 
will contribute and become active. 

Neurons may share into activities of other neurons to the degree that they 
become recruited. This is the idea of neurons being able to support the 
firing of other 
neurons without reading out their own memory. However this may be 
a fairly overlayed process, where neurons may be more or 
less identifiable in this process of read-out, rather than being either 
mere uniform processors or highly individual sources of specific information.

When a neuron is highly activated for "read-out", how can it 
have a distinct signature of contribution to the network 
activation? 
Synaptic connections are often not stable, permament connections,
but undergo significant presynaptic gating by activation of presynaptic 
receptors and intrinsic calcium transients.
The concerted gating of many synapses could have a network-wide 
effect in an overall fast modulation of network connectivity. 
This could support selective read-out from a small cluster of 
neurons followed by spreading the activation in the network
by an increase in overall coupling. 

\section{Implications for memory}
An interesting consequence of this mechanism is that it 
explains the usually latent nature of most memories.
The main mechanism to make fleeting alterations permanent is 
to fixate changes in a time-layered process which usually requires additional 
conditions, to keep changes from flipping back \cite{Scheler2003}.

One way to hide memories in a brain is to make them temporally deep,
so that they are fixed and permanent and are not easily accessed. The 
other way is to make them spatially unconnected. A fairly easy 
way of how this can be done regards the distribution of presynaptic 
and postsynaptic receptors. If a set of neurons is surrounded by 
connections that are all highly conditionally gated in permanent 
settings, then it will be virtually unretrievable.

Specifically, intrinsic plasticity may be explored in relation to the 
extremes in acessibility of memorization,
e.g. associated with traumatic 
experiences, memory suppression and retrieval by specific triggers or 
even brain states.

In those cases there may be isles and pockets of memory that exist fairly 
without contact with any others. These will then require special circumstances 
to become active, possibly also because there are chains of memory events 
involved which require 
several stages and time-frames before they are fully present.

\section{Conclusion}
One wonders whether the physiological conditions of inducing 
synaptic plasticity exclusively by synaptic stimulation truly exist, 
or whether the process of remodeling at a synapse by pre- and postsynaptic 
stimulation does not always require a fairly large amount of concurrent 
events, spaced in time.

Important observations concerning memorization are the 
time-layering of neural plasticity 
mediated by additional constraints to enter into later stages, 
the various manifestations of intrinsic neural properties, and the 
existence of conditional gates on synaptic connections.

The main problem about replacing LTP as a memory mechanism has been 
to find other highly abstract, easily understandable principles for 
induced plasticity.

In this paper we have attempted to lay out such a basic mechanism. 
Besides the observations on intrinsic modulation of the transfer function,
and gated synapses, the idea of potentially compensatory mechanisms of 
storing memories by permament or gated connections, permanent or 
triggerable membrane properties, and hidden properties that require 
read-out to affect neuronal information transfer would require a closer 
look.


\begin{thebibliography}{10}

\bibitem{Burchett2003}
Scott~A Burchett.
\newblock {In through the out door: nuclear localization of the regulators of G
  protein signaling.}
\newblock {\em J Neurochem}, 87:551--9, 2003.

\bibitem{ChaoNestler2004}
Jennifer Chao and Eric~J Nestler.
\newblock {Molecular neurobiology of drug addiction.}
\newblock {\em Annu Rev Med}, 55:113--32, 2004.

\bibitem{EkstromADetal2003}
Arne~D Ekstrom, Michael~J Kahana, Jeremy~B Caplan, Tony~A Fields, Eve~A Isham,
  Ehren~L Newman, and Itzhak Fried.
\newblock {Cellular networks underlying human spatial navigation.}
\newblock {\em Nature}, 425:184--8, 2003.

\bibitem{GarabedianCEetal2003}
Catherine~E Garabedian, Stephanie~R Jones, Michael~M Merzenich, Anders Dale,
  and Christopher~I Moore.
\newblock {Band-pass response properties of rat SI neurons.}
\newblock {\em J Neurophysiol}, 90:1379--91, 2003.

\bibitem{GruberAJetal2003}
Aaron~J Gruber, Sara~A Solla, D~James Surmeier, and James~C Houk.
\newblock {Modulation of striatal single units by expected reward: a spiny
  neuron model displaying dopamine-induced bistability.}
\newblock {\em J Neurophysiol}, 90:1095--114, 2003.

\bibitem{HaasWhite2002}
Julie~S Haas and John~A White.
\newblock {Frequency selectivity of layer II stellate cells in the medial
  entorhinal cortex.}
\newblock {\em J Neurophysiol}, 88:2422--9, 2002.

\bibitem{HampsonREetal2004}
R~E Hampson, T~P Pons, T~R Stanford, and S~A Deadwyler.
\newblock {Categorization in the monkey hippocampus: A possible mechanism for
  encoding information into memory.}
\newblock {\em Proc Natl Acad Sci U S A}, 88:2422--9, 2004.

\bibitem{KohenRetal2003}
Ruth Kohen, John~F Neumaier, Mark~W Hamblin, and Emmeline Edwards.
\newblock {Congenitally learned helpless rats show abnormalities in
  intracellular signaling.}
\newblock {\em Biol Psychiatry}, 53:520--9, 2003.

\bibitem{Scheler2003}
G~Scheler.
\newblock {Regulation of Neuromodulator Receptor Efficacy - Implications for
  Whole-Neuron and Synaptic Plasticity}.
\newblock {\em accepted at Progress in Neurobiology}, 2004.

\bibitem{UslanerJetal2001}
J~Uslaner, A~Badiani, C~S Norton, H~E Day, S~J Watson, H~Akil, and T~E
  Robinson.
\newblock {Amphetamine and cocaine induce different patterns of c-fos mRNA
  expression in the striatum and subthalamic nucleus depending on environmental
  context.}
\newblock {\em Eur J Neurosci}, 13:1977--83, 2001.

\bibitem{WermeMetal2002}
Martin Werme, Chad Messer, Lars Olson, Lauren Gilden, Peter Thoren, Eric~J
  Nestler, and Stefan Brene.
\newblock {Delta FosB regulates wheel running.}
\newblock {\em J Neurosci}, 22:8133--8, 2002.

\end{thebibliography}
\end{document}